\documentclass[showpacs,preprintnumbers,amsmath,amssymb]{revtex4}
\usepackage{graphicx}

\begin{document}
\title{ Unbounded autocatalytic growth on diffusive substrate: the extinction transition}
\author{Sasi Moalem and Nadav M. Shnerb}
\affiliation{Department of Physics, Bar-Ilan University, Ramat-Gan
52900 Israel}
\begin{abstract}
The effect of diffusively correlated spatial fluctuations on the
proliferation-extinction transition of autocatalytic agents is
investigated numerically. Reactants  adaptation to  spatio-temporal
active regions is shown to lead to proliferation even if the mean
field rate equations predict extinction, in agreement with previous
theoretical predictions. While in the proliferation phase the system
admits a typical time scale that dictates the exponential growth,
the extinction times distribution obeys a power law  at the
parameter region considered.
\end{abstract}

 \maketitle
\section{Introduction}
Catalyst induced proliferation systems are very common in a variety
of fields in nature such as biology, chemistry, physics, and even in
finance and social sciences. Technically, there is one type of
agents that catalyze a reaction among other species, like enzymes in
living animals, catalysts in chemical reactions  or traders in
social networks. In many cases the catalyst is not influenced by the
reaction it facilitates. The spatio temporal dynamics of the
catalysts agents dictates the dynamics of the reactive system. The
theoretical understanding of such processes is usually based on
partial differential equations for population \emph{densities}, like
the rate equations or (if the system is not well mixed)
reaction-diffusion equations. Although very useful, these equations
neglect the intrinsic noise build up in any realistic system due to
the stochastic motion of individual reactants. Intuitively, the
approximation involved in the study of deterministic processes for
continuum variables rather then the noisy wandering of discrete
agents is valid at high densities, fast diffusion limit.
Unfortunately, no general systematic methodology for consistent
perturbative expansion around this "mean field" limit is available,
and the corrections are calculated on the basis of an ad-hoc methods
for spacial cases \cite{derrida} or by numerical simulations
\cite{kaneko}.

In a model presented recently \cite{AB,ABpre} the disagreement
between the deterministic rate equations and the realistic
stochastic process is emphasized for a very simple and generic
system. The model includes two species: an immortal  catalyst A that
only diffuses randomly in space, and a reactant agent B which decays
at rate $\mu$ and proliferate in the presence of A-s at rate
$\lambda N_A$, where $N_A$ is the number of A agents (local density
of A) at the reactant spatial location. Both A and B undergo
diffusion with rates $D_A$ and $D_B$ respectively. Schematically,
the local reactions considered are:
\begin{eqnarray}
B &\longrightarrow& \oslash
\\ \nonumber B+A &\longrightarrow& B+B+A.
\end{eqnarray}
The continuum description for this process is given by the
mean-field rate equations for the densities of A and B, $a(x,t)$ and
$b(x,t)$, respectively. The  corresponding reaction diffusion
equations,
\begin{eqnarray}
\frac{\partial a(x,t)}{\partial t}&=&{D_A\nabla^2 a(x,t)} \nonumber
\\\frac{\partial b(x,t)}{\partial t}&=&{D_B\nabla^2 b(x,t)-\left[ \mu-\lambda a(x,t) \right] b(x,t)},
\label{rateEq}
\end{eqnarray}
admit a simple solution: since A only diffuses, catalyst density
becomes a constant $n_A$ (where $n_A$ is the average A density)
after long time, and the second rate equation becomes linear:
\begin{equation}
\frac{\partial b(x,t)}{\partial t}= {D_B\nabla^2 b(x,t)-m b(x,t)},
\label{MF}
\end{equation}
where $m \equiv \mu-\lambda n_A$ is the decay/growth rate of the
system, depending on its sign. This mean field theory predicts a
phase transition at $m=0$. For positive $m$ the reactant
concentration decays exponentially, while negative $m$ yields
exponential growth (proliferation). In realistic systems, of course,
one expects some saturation mechanism that prevents explosion,
perhaps in the form of B agents competition for resources. If  this
work this process is neglected and the system admits two fixed
points, $b=0$ and $b=\infty$. The significance  and limitations of
this approximation are discussed in the last section.

In previous work \cite{AB,ABpre} the effects of stochasticity due to
the discrete character of the reactants was considered for this "AB
model". The main observation is that diffusion never kills spatial
fluctuations for discrete A agents: instead, one gets at any time
Poissonian fluctuations of the A density, entailing  that at any
time there will be a finite chance for a site to be "active" (with
local $m(x) = \mu - \lambda N_A(x)$ smaller than zero) even if the
average $m$ is positive. At each of these "oases" the B population
grows exponentially and a colony of B agents is developed. This
leads to a shift of the transition from $m=0$ to finite, positive
$m$.

\begin{figure}
  \includegraphics[width=7.7cm]{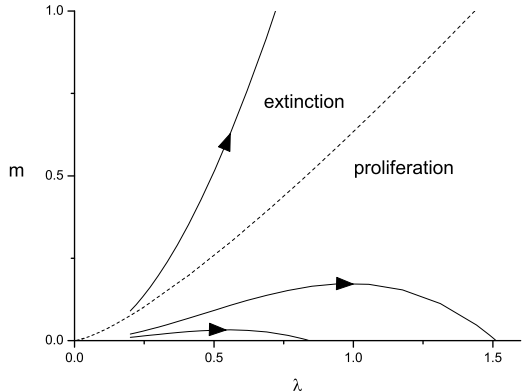}\\
  \caption{RG results for AB model phase transition, $D_A=D_B=2$, $d=1$,
  $n_A=1$ and
  $\Lambda=2\pi/a_0$ where $a_0=1$  is the lattice constant.
  Above the dashed line is the extinction phase for infinite system while below is the proliferation region where the
  death rate $m$ flows to negative values.
  One can see that catalyst fluctuation induce proliferation in area where the  mean-field predicts extinction
  ($m>0$).
  Finite system size implies truncation of the  RG flow lines, so the final effective  $m$ value may
  be positive even if the bare mass is in the proliferation region.   }
  \label{RG_PhaseTrans}
\end{figure}

Let us present the main results  of \cite{AB,ABpre}.  The technical
tool used  is a general approach for the consideration of the effect
of stochastic noise in reactive systems on spatial domains,
suggested and applied by various authors \cite{Doi,Peliti,Cardy}.
This theoretical framework is  based on the exact Master equation of
the full stochastic process. Translating the Master equation to the
form of Schr\"{o}dinger like equation for second quantized fields
one may find the effective action and integrate fast modes to get
the same effective action with reaction parameters that depend on
time and length scale (renormalization group flows). Here, the RG
transformation involves the rescaling:

\begin{eqnarray}
x&\rightarrow& sx, \nonumber
\\t&\rightarrow& s^zt,  \nonumber
\\a&\rightarrow& s^{-d}a,
\\b&\rightarrow& s^{-d}b,  \nonumber
\\\Lambda&\rightarrow& \Lambda/s, \nonumber
\label{trans}
\end{eqnarray}
where $s$ is the scaling factor ($s>1$), $d$ is the dimension and
$\Lambda$ is the upper momentum cutoff. This rescaling yields the
flow equations for the model parameters:

\begin{eqnarray}\label{RG}
\frac{d \lambda}{d \ln s}&=&(2-d)\lambda+\frac{\lambda^2}{2\pi D}
\frac{\Lambda^{d-2}}{1+\frac{m}{D \Lambda^2}} \nonumber
\\\frac{d m}{d \ln s}&=&2m-\frac{n_A\lambda^2}{2\pi D} \frac{\Lambda^{d-2}}{1+\frac{m}{D\Lambda^2}}
\\\frac{d n_A}{d\ln s}&=&d n_A, \nonumber
\end{eqnarray}
where the average diffusion rate is $D=(D_a+D_b)/2$. These flow
equations were obtained by one loop perturbative expansion of the
effective action around its mean field limit, where $\lambda/D$ and
$m/D$ are the small parameters. One can see that there is a negative
correction to $m$ and  this implies that the effective "mass" $m$
may flow to the proliferation phase $m<0$ as the system size
increased, even if the bare mass is larger than zero. A sketch of
the flow lines according to Eqs. (\ref{RG}) is presented in Figure
(\ref{RG_PhaseTrans}).

The aim of this work is twofold: first, to compare the theoretical
predictions about the location of the extinction transition  with
numerical simulations; second, to study the nature of that phase
transition. Clearly, the existence of proliferation phase with
diverging number of reactants compels a simulation that involves
with large number of agents, and is possible only when $D$ is
comparable with the reaction parameters. Accordingly, the RG results
should be taken only as a qualitative guides, as the "small
parameter" of the expansion is $O(1)$ in the numerical experiments.
In Figure (\ref{RG_PhaseTrans}) these theoretical predictions for
the extinction-proliferation phase transition, along with few RG
flow lines are  shown  for $d=1$, $n_A=1$, $D=2$ and
$\Lambda=2\pi/a_0$ where $a_0=1$  is the lattice constant.

The intuition gathered from the RG approach is that there are two
parameter regions. In the proliferation phase the effective coupling
constant $\lambda$ grows with the  basic length scale of the system
and for infinite sample this process leads to negative $m$. In the
extinction phase, on the other hand, the growth of $m$ dictates the
system and the flow is toward extinction.

In order to implement this intuition to finite samples, one should
bear in mind that positive mass corresponds to extinction. Thus, the
crossover to proliferation depends on the ability of the system to
support more and more rescaling transformations: even if the bare
mass is in the  proliferation region of Figure \ref{RG_PhaseTrans},
system of finite size may be in the extinction phase, depending on
the  bare parameters $\lambda$, $m$ and $n_A$.

In the next section the numerical technique is presented, the
location of the transition is identified for finite samples and the
results are compared with the theoretical predictions. The third
section deals with the statistical properties of the results in
different regimes where the essential difference between the phases
is emphasized. In the last section we discuss the results and
compare our study with known facts about extinction transition for
bounded growth like the contact process in homogenous and
heterogenous environments.

\section{ Numerical study of the extinction transition}
\subsection{Simulation technique}

In order to simulate the AB model,  a one  dimensional array of $L$
lattice points  with periodic boundary conditions is used. Each site
$i$ is associated with an integer ($N_A^i$) number of catalysts and
another integer ($N_B^i$) stands for the number of reactants. The
hopping rates of the catalyst and the reactants are $D_A$ and $D_B$
respectively. The death rate of a reactant is $\mu$, while the
proliferation rate at each site is given by $\lambda N_A^i$. The
initial densities of the catalyst and the reactant are $n_A$ and
$n_B$ correspondingly. For each system realization the catalysts are
initially randomly distributed. The mean-field effective mortal rate
(also defined as the reactant "mass" in the corresponding effective
field theory) is therefore $m=\mu-\lambda n_A$.

The main technical obstacle for a single agent based simulation of
that system is the large number of reactants in the proliferation
phase. It turns out that an event driven algorithm based on
individual agents  leads to diverging simulation times way before
any conclusive statement about the actual phase of the system is
extracted. In particular, even a case of very large reactant
"colony" may be recognized as a transient that ends up with complete
extinction of the colony. Accordingly, a site based simulation has
been used. For heavy populated sites ($N_B^i>{10^{6}}$) the
mean-field equations (\ref{MF}) are solved with  gaussian noise
proportional to the square root of the reactant density,  while if
$N_B^i<10^6$ a "global" Monte Carlo method is used.  For example,
the number of particles that decay ("died") at certain site per
small unit time $dt$ is determined by generating a random number
taken from a binomial distribution with probability $\mu dt$ and
$N_B^i$ Bernoulli trials (instead of checking $N_B^i$ times whether
a reactant died during this time step). This combination of two
methods allow us to reach a huge number of reactants, up to
$N_B=10^{300}$, without neglecting the effect of discretization at
low densities.

The  simulation is initiated with reactant density  $n_B=100$ and
catalyst density $n_A=1$.  All the results presented here correspond
to reactant and catalyst diffusion rates $D_B=D_A=2$. The "mass" is
therefore $m=\mu-\lambda n_A=\mu-\lambda$. The simulation of an
individual process  ends when the system reaches either extinction
or proliferation.

An extinction of the process is declared if there are no reactants
in the system. The definition of proliferation phase, on the other
hand, is more subtle as  there is no upper bound for the reactant
density. We thus set an upper threshold of $N_B=10^{50}$; above this
number the simulation stops and the phase of the system is
identified as proliferation phase. This criterion was justified by
increasing the upper bound to be so large that transient effects are
avoided. In fact, setting the threshold to be $N_B=10^{300}$ no
significant differences were identified.

\subsection{The transition: theory vs. simulations}

\begin{figure}
  \includegraphics[width=7.7cm]{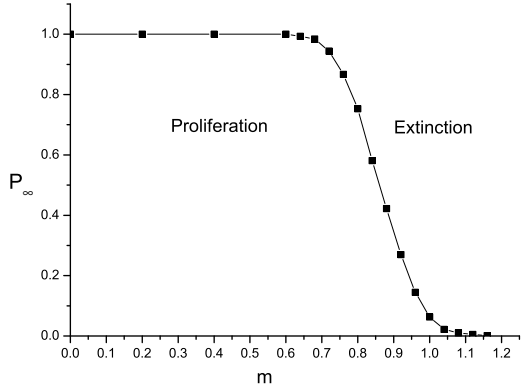}\\
  \caption{Survival probability (average over 1,000 realizations) $P_\infty$ vs $m$,  $\lambda=1$,
  $n_A=1$, $n_B=100$.
  According to the reaction  diffusion equations
  the extinction phase is above $m=0$.  Here one finds a region of almost sure proliferation and
a region of almost sure extinction, separated by an uncertain
region. The proliferation region grows with the system size $L$ and
saturates at a critical value, $m_c$, that correspond to the
extinction transition for an infinite system.  }
  \label{fig1}
\end{figure}

As the system is stochastic in nature, the flow to the extinction or
proliferation phase is non-deterministic, and  for the same set of
parameters one may find microscopic realizations that leads to
different states. In Figure (\ref{fig1}), for example, the
probability $P_\infty$ for the system to reach the proliferation
phase is plotted against $m$ for $L=128$ and $\lambda =1$, where
each point reflects the fraction of systems (out of  1000 runs) that
reach proliferation. According to the mean field theory (\ref{MF})
above $m=0$ the system should flow to the extinction phase. In
contrary, here one finds a region of almost sure proliferation and a
region of almost sure extinction, separated by an uncertain region.
In what follows (for the sake of comparison between the numerics and
the analytic prediction for the location of the phase transition)
the definition of the extinction transition is taken at $m$ for
which $P_\infty = 0.5$, in order avoid the stronger effects of noise
as $P_\infty$ approaches zero or one.

\begin{figure}
  \includegraphics[width=7.7cm]{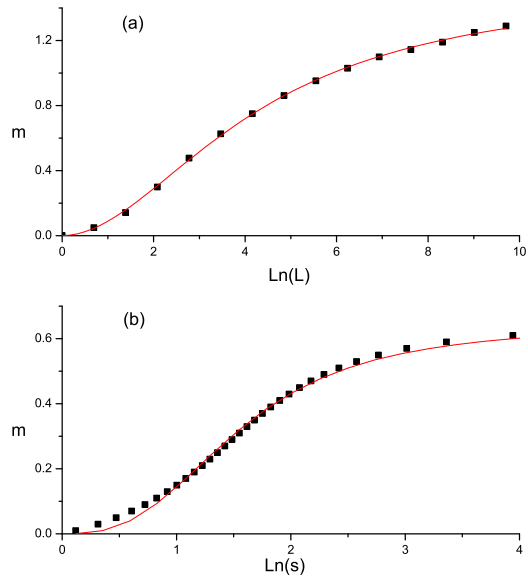}\\
  \caption{$m$ Vs $\ln(L)$: (a) Simulation results (squares)  for   $\lambda=1$, $n_A=1$, $n_B=100$.
  One can see that the system robustness is increasing (proliferation for larger death rates)
  as the system size grows.
  The critical "mass" $m_c$ (the mass for which there is no proliferation even in the limit $L \to
  \infty$)
  is determined using the best fit to the logistic
  function (\ref{fit}) (full line).
  (b) RG results (squares) for  $\lambda=1$, $n_A=1$ and
  $\Lambda=2\pi/a_0$ where $a_0=1$ is the lattice constant. There is a
  region where the phase transition
  depends on the system size, but above a critical mass $m_c$
  there is no growth independent of $L$, in agreement with Figure
  (\ref{RG_PhaseTrans}). The full line is the best fit to  (\ref{fit}).
   The qualitative resemblance between the RG predictions to the
simulation results is clear.}
  \label{m_Vs_L}
\end{figure}

Using that analysis it is possible to  continue our quest of finding
the critical mass $m_c$ for the (infinite size system)
extinction-proliferation phase transition. As a first step, the
system size $L$ is multiplied  by 2 while the reaction rates and
agents densities  are kept constant. Next the survival probability
$P_\infty$  is plotted vs $m$ for  each system size (in order to
change $m$  only $\mu$  is modified while  $\lambda$ and $n_A$
 are kept constants). As mentioned, the phase transition
 for a finite size system
is defined where $P_\infty=0.5$. The critical $m$ of the finite size
system  is plotted versus $\ln L$ in  Figure (\ref{m_Vs_L}.a). One
can see that for small system sizes, the critical "mass" increases
as the system size $L$ increases, since the probability for larger
catalyst fluctuation grows with size. There is, however, an
intrinsic limit (independent of the system size) on these spatial
fluctuations, and according to the RG analysis one expects that for
large enough bare mass there will be no further increase in the
critical mass as a function of the system size [see Figure
(\ref{m_Vs_L}.b)]. Indeed, the simulation indicates that for larger
system sizes the critical mass begins to saturate while it
approaches the "real" phase transition.  Since the "real" phase
transition at the thermodynamic limit is beyond  our computational
abilities, the critical mass $m_c$ for certain $\lambda$ is
evaluated by fitting the results to the function
\begin{equation}\label{fit}
    m(x) = m_c -\frac{m_c}{1+(x/x_0)^p}
\end{equation}
where $x \equiv ln(L)$. As indicated in Figure (\ref{m_Vs_L}.b),
this function yields an almost perfect fit to the RG based theory,
and seems also to fit the numerical experiment quite well [Figure
(\ref{m_Vs_L}.a)].

\begin{figure}
  \includegraphics[width=7.7cm]{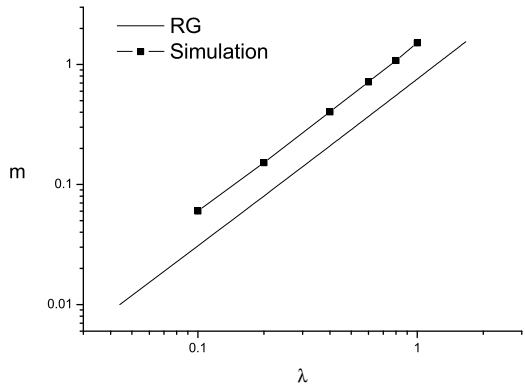}\\
  \caption{Phase transition lines in the  $m$ - $\lambda$
   plane. The full line is the theoretical predication based on the RG
   perturbative analysis while the squares are the results of the numerical experiment
   (see text).}
  \label{mVsG}
\end{figure}

After  the critical mass for a specific $\lambda$ is determined,
using the above procedure,  one proceeds to plot the infinite system
size phase transition line $m_c$ vs $\lambda$, Figure (\ref{mVsG}).
Above the transition line is the extinction phase for the infinite
size system, while below is the proliferation phase. The theoretical
phase transition line, based on the RG analysis, is plotted together
with the line obtained from the numerical experiment on a
logarithmic scale. Both theoretical and numerical results suggests
that $m_c \sim \lambda^{\alpha}$, where the values of $\alpha
\approx 1.4$ (for the RG phase transition curve) and $\alpha \approx
1.45$ (for the numerics) are close and the curves are almost
parallel. This, again, implies a qualitative similarity between the
numerical simulations and the picture emerges from the theory.

\section{Sample to sample fluctuations and the phase transition}

\begin{figure}
  \includegraphics[width=7.7cm]{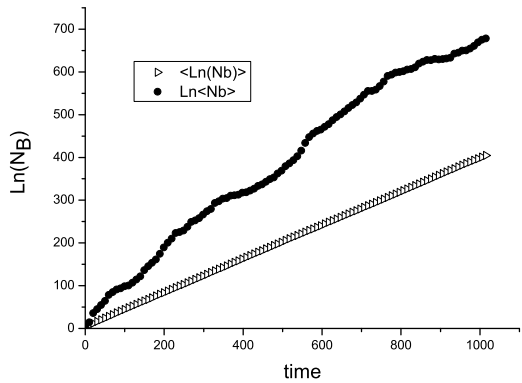}\\
  \caption{$\ln\langle N_B \rangle$ and $\langle\ln(N_B)\rangle$ vs time (average over 1,000 realizations),
   $\lambda=1$, $m=0.6$, $n_A=1$, $n_B=100$. The typical case is shown to be similar to the
  average over all realizations. The lines do not collapse since $\langle\ln(N_B)\rangle$ is dominated by the peak
  of the gaussian of Figure (\ref{fig3}), while $\ln\langle N_B \rangle$ is dominated the tail of this histogram.}
  \label{fig2}
\end{figure}

\begin{figure}
  \includegraphics[width=7.7cm]{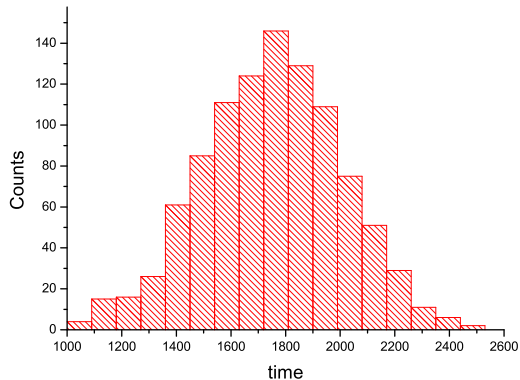}\\
  \caption{End times histogram (1,000 realizations),
   $\lambda=1$, $m=0.6$, $n_A=1$, $n_B  =100$. One can see that there is a typical time for the
system to reach the proliferation upper bound.}
  \label{fig3}
\end{figure}

Let us take  a deeper look at the dynamics of the system, trying to
describe the different regions of Figure (\ref{fig1}). For a point
in the distinguished proliferation region ($m < 0.6, P_\infty \sim
1$), all the samples proliferate. The growth of the B population is
exponential, and both the average over the logarithm of the
population and the logarithm of the average grow linearly in time
[Figure (\ref{fig2})]. (if most of the realization decay and only
few show exponential growth, $\langle N_B \rangle$ is dominated by
rare events and grow, while $\langle \ln(N_B) \rangle$ decays to
zero since the logarithm kills the influence of rare events on the
average.) If we further plot a histogram of the "end times" for
various realization (i.e., the time needed to the system to reach
the proliferation upper bound) one can see that there is a typical
end time and the histogram corresponds to Gaussian distribution
around its value [Figure (\ref{fig3})]. The lines of Figure
(\ref{fig2}) do not coincide since $\langle\ln(N_B)\rangle$ is
dominated by the peak of the gaussian, while $\ln\langle N_B
\rangle$ is dominated by its tail.

\begin{figure}
  \includegraphics[width=7.7cm]{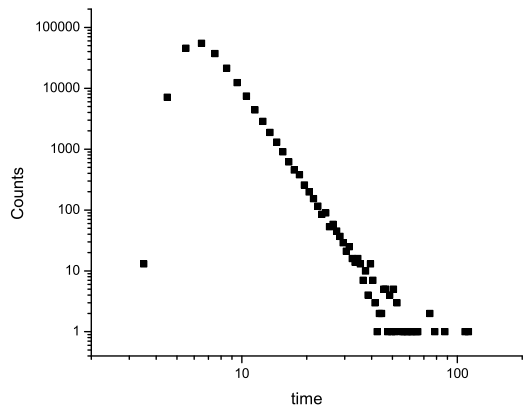}\\
  \caption{Death times histogram (log-log scale, 200,000
  realizations) for
   $\lambda=1$, $m=2$, $n_A=1$, $n_B=100$. The death times histogram seems to maintain its
  power-law  tail even above the "real" extinction phase transition. However,  it decay much faster
   with exponent $\theta  \sim 5$. The rare events in that case are "too rare" and the average is dominated by the
   typical events, as emphasized in Figure \ref{fig7}. }
  \label{fig6}
\end{figure}

\begin{figure}
  \includegraphics[width=7.7cm]{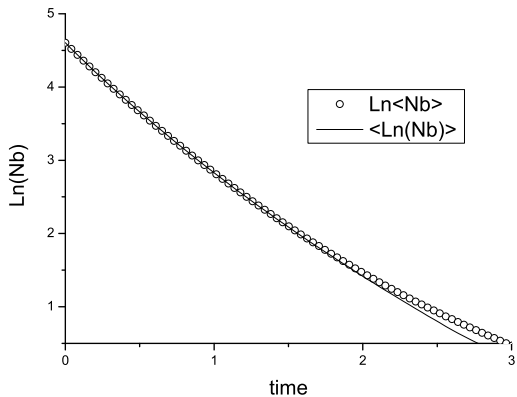}\\
  \caption{$\ln\langle N_B \rangle$ and $\langle\ln(N_B)\rangle$ vs time (average over 1,000 realizations),
    $\lambda=1$, $m=2$, $n_A=1$, $n_B=100$. One can see that the average case is similar to the
  average over the realizations. There are no significant fluctuation (within our finite realization number)
  since finding exponentially rare events is almost impossible}
  \label{fig7}
\end{figure}

Deep in  the extinction region (e.g.,  $m=2, P_\infty=0$) the
situation is different.  In this region all the samples end up in
extinction, but the corresponding time scale (the effective decay
rate, or the effective mass) shows strong sample to sample
fluctuations. In Figure (\ref{fig6}) a histogram of death times (the
inverse mass) is shown. If one assumes that, at long times, the
reactant concentration decays like $\exp(-t/\tau)$, this figure
implies that $P(\tau) \sim \tau^{-\theta}$ where for the results of
figure (\ref{fig6}) $\theta \approx 5$. The strong decay of $P(t)$
manifest itself in the similarity between the typical and the
average cases, Figure (\ref{fig7}).

\begin{figure}
  \includegraphics[width=7.7cm]{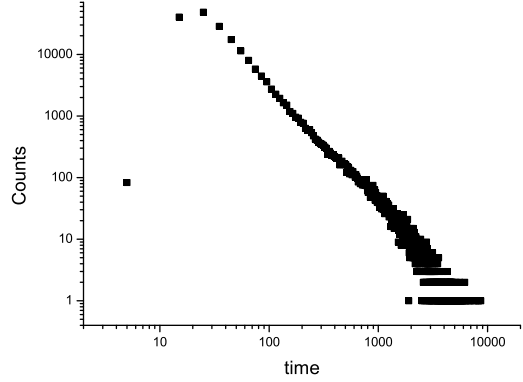}\\
  \caption{Log-log scale death times histogram (200,000 realizations),
   $\lambda=1$, $m=1.1$, $n_A=1$, $n_B=100$. The probability of realization to decay ($N_B = 0$) between
  $t$ to $t+\Delta t$ is shown.  The data correspond to a  power law distribution of decay times with
   $P(\tau) \sim \tau^{-\theta}$ with $\theta \approx 2$. This implies
  that, to the accuracy of our numerical experiment,
the average time  associated with the decay of the system in the
extinction phase diverges.}
  \label{Hist_m=1.1}
\end{figure}

\begin{figure}
  \includegraphics[width=7.7cm]{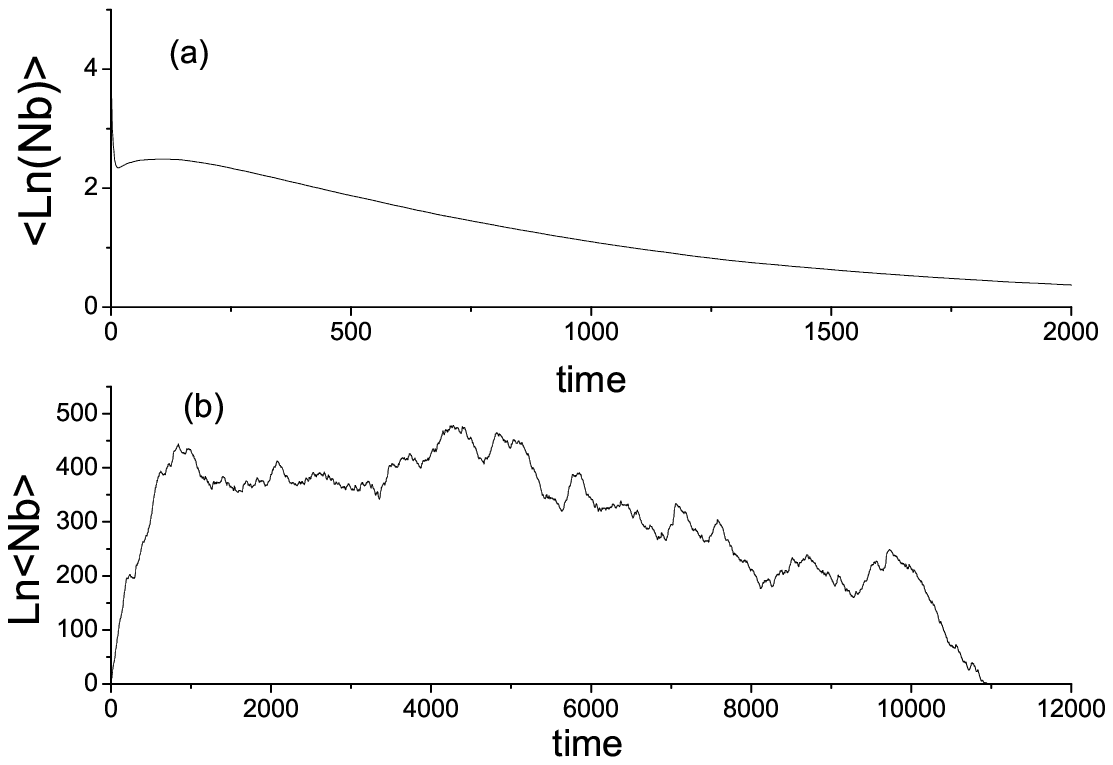}\\
  \caption{ Typical (a) versus average (b) decay patterns. In the upper panel  $\langle\ln(N_B)\rangle$ is plotted
  against time and shows a smooth decay, while in the lower panel   $\ln\langle N_B \rangle$ shows completely different
  pattern with different time scales. The average is taken from   200,000 different
  realizations,for
  $\lambda=1$, $m=1.1$, $n_A=1$, $n_B=100$.
  Most of the samples decay to extinction very fast, but the
  average (b) is dominated by rare events, due to strong sample to
  sample fluctuations.}
  \label{AvrLn_m=1.1}
\end{figure}

Close to  the edge of proliferation (e.g.,  $m=1.1, P_\infty << 1$)
the power law for the decay is much smaller. In Figure
(\ref{Hist_m=1.1}) the histogram of death times indicates $\theta
\approx 2$.
 Accordingly,  there is a lot of difference
between the average ($\ln \langle N_B \rangle$ ,Figure
(\ref{AvrLn_m=1.1}.a)) and the typical ($\langle \ln(N_B)\rangle$,
Figure (\ref{AvrLn_m=1.1}.b)) time evolution of the total B
population. While the typical case decays smoothly to zero, on
average the system  grows, saturates, and decays after much larger
time. The samples for which $\tau$ is large are samples where the
reactant population grows rapidly on time scales smaller than $\tau$
and then decays. Accordingly, these samples dominate the average
population while the typical population is determined by the large
number of samples that admits small $\tau$ \cite{kesten1}.

\begin{figure}
  \includegraphics[width=7.7cm]{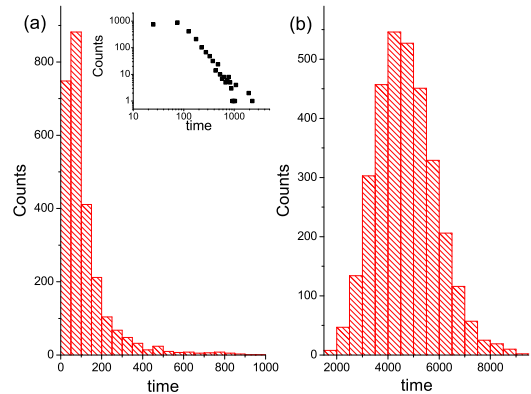}\\
  \caption{Histogram of death times (a) and of proliferation times (b) in the intermediate parameter region.
  One sees that the  death times distribution admits
  a long tail corresponds to power-low [see the inset of panel (a)],
   while proliferation times distribution is sharply peaked around its typical value.}
  \label{m=0.85}
\end{figure}

In the middle region, say, $0.6<m<1.1$,  the survival probability is
between zero and one, and accordingly  two peaks of death and
proliferation times are seen in Figure (\ref{m=0.85}), corresponding
to the proliferating  and to the extinction events. Again there is a
Gaussian like distribution around a typical time scale associated
with the proliferation samples, and power law distribution of time
scales in the extinction samples.

Our numerical experiment suggest, thus, that the extinction
transition of the saturation free AB model takes place in the
following way: deep inside the extinction phase there is a power law
distribution of decay times with $\theta
> 2$, and $\theta$ becomes smaller and smaller while decreasing $m$.
Close to the proliferation threshold   $\theta \sim 2$, and from now
on there is finite fraction of the samples that proliferate to
infinity. The sample to sample fluctuations decrease as the
transition is approached, and in the "living" phase Gaussian
statistics is observed.

\section{conclusions and discussion}

This paper presents a numerical experiment that confirms,
qualitatively, some of the theoretical predictions about the
extinction transition in system of autocatalytic agents with
diffusively correlated fluctuations. The extensive simulations used
render a phase diagram similar to the predictions of the exact
Master equation based  renormalization  group technique, even in the
parameter region where the "small parameters" of the perturbative
technique are $O(1)$.  As expected, resources fluctuations increase
the ability of species to survive, but in order to extract the
potential of these fluctuation via the adaptation mechanism that
associate reactant colonies with catalysts islands, the size of the
system must be large enough. This finite size effect has been
demonstrated, and the system is taken to the large scale limit in
order to approach the "real" phase transition.

Another result, not predicted by the theory, is the appearance of a
scale free fluctuations in the extinction phase, in contrary with
the Gaussian distribution associated with the characteristic time in
the growth phase.

Let us compare, now, these findings with some known results about
extinction transitions. A common conjecture, first suggested by
Grassberger and Janssen \cite{grassberger,janssen} is that, in the
absence of conservation laws, all the extinction transitions belong
to the directed percolation (DP) equivalence class
\cite{hinrichsen}. This proposal is based on the following
intuition: close to the transition point the average density of the
reactants is zero. In a single time step reactant may decay, survive
or yield an offspring. These processes are the basic rules of DP,
hence all transitions are at the same equivalence class.

In the presence of spatial heterogeneity, on the other hand, the
situation is more complicated as quenched randomness appears to be a
relevant operator. It was shown by Janssen \cite{janssen1} that the
corresponding field theory admits only runaway solutions and no
stable fixed point is reached in the physical domain, so the
transition in that case is not in the DP equivalence class. This
analysis  supports  the  numerical work  of Moreira and Dickman
\cite{moreira}, who used Monte Carlo simulations to analyze the 2d
contact process with various fractions of randomly  distributed
inactive sites.  In the extinction region these authors identify the
Griffith phase, with  nonuniversal power laws. In that regime the
\emph{local} dynamics, associated with rare events,  dictates the
system's evolution. While the chance to find a "good" region (with
no inactive sites) of linear size $R$ decays exponentially with $R$,
its typical lifetime \emph{grows} exponentially with size, yielding
a non universal power law below the transition.

One may a-priori suggest that, for the AB model considered here,
the diffusive dynamics of the catalysts simply average out their
effect, thus the resulting transition should be in the directed
percolation equivalence class, like the transition on homogenous
substrate. This assumption is wrong, since a diffusive heterogeneity
is a relevant operator at the transition. One may either apply the
Harris criterion or consider the naive scaling for the  action
associated with the Reggeon field theory with diffusive disorder:
\begin{equation}
S = \int dt d\textbf{x} \ \  \psi^* \frac{d \psi}{dt} + D_{\psi}
(\nabla \psi^*) (\nabla \psi) + u (\psi^* \psi \psi - \psi^* \psi^*
\psi) + \phi^* \frac{d \phi}{dt} + D_{\phi} (\nabla \phi^*)(\nabla
\phi) - v (\phi \psi^* \psi + \phi^* \psi^* \psi)
\end{equation}
to see that it supports only runaway solutions, and that $D_{\phi}$
flows to zero upon rescaling transformation.

Since the system seems to  flows into the quenched disorder phase
one may suggest that, for an unbounded growth, the colonies on the
rare oases grow without limit, so there is no phase transition in
the thermodynamic limit (unless the catalyst distribution and
dynamical rules are such that they  never allows for $m<0$ at any
finite region). However, Kesten and Sidorovicious \cite{kesten}
proofed that, even without saturation term, there is a parameter
regime that corresponds to the extinction phase, where all processes
decay with probability one for an infinite sample. The stochastic
wandering of the catalysts, thus, seems to control the system and to
yield a possibility of extinction even in the infinite capacity
limit.

The emerging picture, supported by the numerical simulations
presented here, is that the phase transition exists, with a wide
(power law) distribution of extinction times (Griffith phase) below
that transition. The diffusive wandering of the catalysts is not
strong enough in order to  replace the fluctuating concentration by
its average and the adaptation of the reactant colonies to the
Poissonian noise is of  crucial importance. On the other hand, even
if the carrying capacity is unbounded this wandering \emph{is}
strong enough to yield a finite range of parameters where the
process ends up at extinction with probability one.

\section{Acknowledgements}
We thank Prof. David Kessler and Prof. Sorin Solomon for very
helpful discussions about both  theory and  simulation techniques.
This work was supported by the Israeli Science Foundation (grant no.
281/03) and by Yeshaya Horowitz Fellowship.


\begin{thebibliography}{}
\bibitem{kesten1} It was already proofed by Kesten \cite{kesten} that  the average number
of reactants approaches infinity (in one and two dimensions) due to
rare events, even if the typical system decays. This phenomena,
however, depends on exponentially rare events that yields super
exponential population growth and apparently  not related to the
power law extiction times presented here.
\end{thebibliography}
\end{document}